\documentclass[10pt,letterpaper]{article}
\usepackage{opex3}
\usepackage{cite}
\usepackage{amsmath}
\usepackage{array}

\begin{document}

\title{Reflectarray antennas for \\terahertz communications}

\author{
Tiaoming Niu$^{1,2}$,
Withawat Withayachumnankul$^1$,\\
Benjamin S.-Y. Ung$^1$,
Hakan Menekse$^3$,
Madhu Bhaskaran$^3$,\\
Sharath Sriram$^3$,
and Christophe Fumeaux$^1$
}

\affil{$^1$School of Electrical $\&$ Electronic Engineering, The University of Adelaide,\\ Adelaide, SA 5005, Australia\\$^2$School of Information Science and Engineering, Lanzhou University, \\Lanzhou 730000, P. R. China\\$^3$Functional Materials and Microsystems Research Group, School of Electrical and \\Computer Engineering, RMIT University, Melbourne, VIC 3001, Australia}

\email{[niutm,withawat,cfumeaux]@eleceng.adelaide.edu.au}




\begin{abstract}
Reflectarrays composed of resonant microstrip gold patches on a dielectric substrate are demonstrated for operation at terahertz frequencies. Based on the relation between the patch size and the reflection phase, a progressive phase distribution is implemented on the patch array to create a reflector able to deflect an incident beam towards a predefined angle off the specular direction. In order to confirm the validity of the design, a set of reflectarrays each with periodically distributed $360\times360$ patch elements are fabricated and measured. The experimental results obtained through terahertz time-domain spectroscopy  (THz-TDS) show that up to nearly $80\% $ of the incident amplitude is deflected into the desired direction at an operation frequency close to 1~THz. The radiation patterns of the reflectarray in TM and TE polarizations are also obtained at different frequencies. This work presents an attractive concept for developing components able to efficiently manipulate terahertz radiation for emerging terahertz communications.
\end{abstract}



\section{Introduction}

The concept of reflectarrays can be dated back to the early 1960s~\cite{first_rf}. Combining the principles of phased arrays and geometrical optics, a reflectarray can produce predesigned radiation characteristics without requiring a complicated feeding network. This operation can be achieved by using an array of passive elements, whose individual reflection phase is dependent on a critical dimension of a resonant structure~\cite{book_reflectarray}. To some extent, the performance of the reflectarray is mainly based on the maximum range of the phase change that can be achieved through optimization of the single elements. Normally, the expectation for constructing a reflectarray is that the possible phase change for the single element can cover one $360^\circ$ cycle. This is sufficient for narrowband operation, as the phase can be wrapped to attain larger phase variations. Owing to the operation simplicity, various reflectarrays have drawn intensive interest in the last few decades~\cite{montgomery1978microstrip,pozar1993analysis,gianvittorio2006reconfigurable,chang1995multiple,infrared_rflarray,patent}.

A large variety of resonant elements have been employed to achieve the desired reflection phase change with a dependency on one of their characteristic dimensions. For instance, the reflection phase from a stub-loaded metal patch element is varied by changing the length of the attached stub~\cite{javor1995design}. Further to that, the phase response of a microstrip element can be tuned by varying the size of the metal patch\cite{Encinar_micro_rflarray}. Some more sophisticated structures include the ``phoenix cell'' with rebirth capability that provides nearly $360^\circ$ phase change~\cite{moustafa2011phoenix}, and multilayered structures that provide an alternative for increasing the bandwidth of operation, however, at an expense of the simplicity~\cite{encinar1996design,encinar2001design}.

As for the operation frequency, various reflectarray structures have been intensively realized in the microwave region. Since metals perform like nearly ideal conductors in the microwave band, metal patches with variable dimensions are common for building up reflectarrays, and can reflect the incident waves with high efficiency\cite{encinar2010recent}. As an example of successful applications, a 1.2-meter reflectarray antenna made of three stacked layers containing varying-sized patches has been demonstrated to satisfy the demanding requirements of satellite communication~\cite{Encinar_micro_rflarray}. The structure works in two separate frequency bands of $11.7-12.2$~GHz and
$13.75-14.25$~GHz. Meanwhile, Hu~\emph{et~al.}~\cite{hu2008design} proposed a millimeter wave reflectarray with phase agile elements consisting of identical microstrip patches and a liquid crystal layer over the ground plane. By applying two extremal bias voltages to the liquid crystal, the wide range of the phase change can be obtained with reasonable loss at both 102~GHz and 130~GHz. Beyond the millimeter wave range, the concept of reflectarray has been extended to the infrared band, where a binary phase reflectarray has been realized using subwavelength metallic patches on a dielectric substrate to act as a reflective Fresnel zone plate~\cite{infrared_rflarray}. In addition, nano-sized spherical particles with a core-shell structure have been investigated as concept for an optical reflectarray~\cite{ghadarghadr2009plasmonic,nano_re_array_design}. By independently configuring the material properties or radii of the core and shell structures, the reflected phase change can be controlled. Due to its complexity, the core-shell reflectarray remains a theoretical concept.

For the terahertz spectrum, driven by emerging solid-state sources and detectors, high-gain antennas are required for construction of wireless networking or imaging systems. Low-loss terahertz reflectarray antennas thus promise attractive advantages in manipulating the terahertz radiation. Up to now, however, no reflectarray has been realized for terahertz radiation at around 1~THz and above. But it is certainly worth mentioning some implementations of terahertz \textit{phased} arrays such as the photoconducting antenna array with 64 electrodes by Froberg~\emph{et~al.}~\cite{thz__from_photoconducting} and the $4\times4$ patch antenna array for indoor terahertz communication by Islam~\emph{et~al.}~\cite{thz_array_feed}. The former is sensitive to the electric noise, and the complexity of realization could be a potential limitation of this approach. The latter can become very complicated and is prone to high losses when the number of array elements is large. In addition, Maki~\emph{et~al.}~\cite{maki2009terahertz} demonstrated a terahertz electro-optic source based on the principle of phased array. It was shown that the terahertz beam radiated from the crystal can be steered by controlling the incident angle of the pumped beams without using actual phase shifters. As a component for terahertz communications, an electrically tunable terahertz phased array has been proposed by Monnai~\emph{et~al.}~\cite{monnai2012terahertz}. The array is composed of subwavelength MEM elements that can be reconfigured to scatter and focus surface waves dynamically.

Towards improving the flexibility of controlling the direction of terahertz radiation,  we propose in this paper the realization of reflectarrays operating at the terahertz band. The structures employ square metal patches as resonant phase-controlling elements. Particular attention is paid to the choice of suitable materials at terahertz frequencies, while the tolerances of manufacturing techniques are taken into account. The single element is optimized by simulations employing a Drude model expression for the metal surface impedance, and the relation between the phase response and the patch size is obtained. Based on this relation, a reflectarray is designed to deflect an incident wave into a predesigned angle off the specular direction. The wave deflection capability is essential for terahertz communications to alleviate the line-of-sight limitation~\cite{kleine2011review}. In order to verify the design, the terahertz reflectarrays have been fabricated and the performance of the reflectarrays has been experimentally evaluated by using THz-TDS.

\section{Principle of reflectarray for angular deflection}

\begin{figure}[htbp]
\centering\includegraphics[width=12cm]{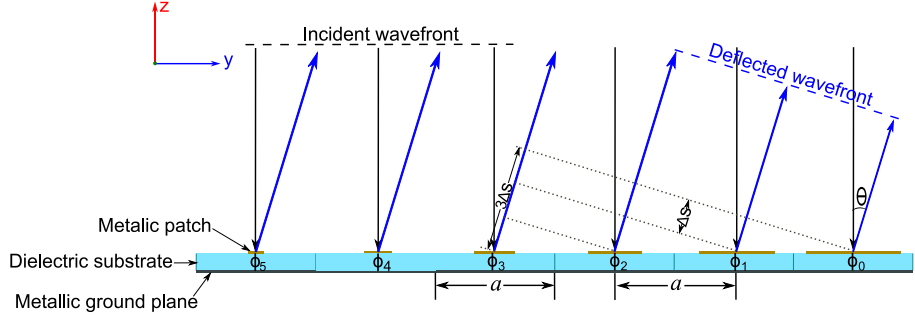}
\caption{Operation principle of the designed reflectarray. The phase distribution results in deflection of a normally incident plane wave towards predesigned angle $\theta$. Here, $a$ indicates the spacing between the center points of two adjacent elements, and $\phi_{i} (i = 0,1,2,3,4,5)$ indicates the phase change introduced by the corresponding element. \label{phase change}}
\end{figure}

The diagram shown in Fig.~\ref{phase change} illustrates the operation principle of a reflectarray for angular deflection. It shows an array of microstrip patch elements with progressive phase changes, designed to steer the reflected beam away from the specular direction. According to the principle of equality of optical paths, it can be computed for this geometry with square resonant patches that the wave incident normal to the surface will be deflected towards an angle of $\theta$, regardless of the polarization of the incident wave. For the $n^{\rm th}(n=0,1,...)$ element introducing the phase change $\phi_{n}$, the following condition must be satisfied

\begin{equation}
\label{eq1}
\phi_{0}+nk_{0}\Delta s=\phi_{n},
\end{equation}
where $k_{0}$ is the propagation constant of the wave in free space, and $\Delta s$ the optical path difference between the $n^{\rm th}$ and $(n+1)^{\rm th}$ elements after reflection. The optical path difference $\Delta s$ can be expressed in the present case in terms of the deflection angle $\theta$ and unit cell dimension $a$, i.e. $\Delta s = a \cdot \sin\theta $. If we define the progressive phase change as $\Delta\phi=\phi_{n+1}-\phi_{n}$,  Eq.~\ref{eq1} can be rewritten as
\begin{equation}
\label{eq2}
 \Delta\phi = k_{0}\Delta s = \frac{2\pi}{\lambda_{0}}\cdot a\cdot\sin\theta,
\end{equation}
or
\begin{equation}
\label{eq3}
\sin\theta = \frac{\Delta\phi\lambda_{0}}{2\pi a}.
\end{equation}
Eqs.~\ref{eq2} and \ref{eq3} describe the inter-dependence between the value of the deflection angle $\theta$ and the progressive phase change $\Delta\phi$. If the deflection angle is specified, the progressive phase change can be calculated with the relation shown in Eq.~\ref{eq2}. For the simplicity of design and fabrication, we adopt here a strategy with periodically arranged identical linear sub-arrays. Therefore, the progressive phase change becomes an integer fraction of $360^\circ$, e.g. a 6-element sub-array will require a progressive phase change of $60^\circ$ to satisfy the periodicity condition. With this strategy, it is possible to determine a specified deflection angle $\theta$ by using Eq.~\ref{eq3}.

\section{Specific design of terahertz reflectarrays}

\subsection{Patch element and its characteristics}

\begin{figure}[htbp]
\centerline{\includegraphics[width=7cm]{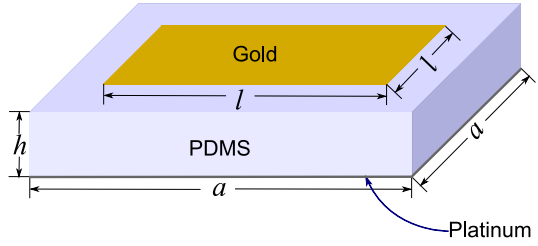}} \caption{A unit cell for the reflectarrays with $a=140 \hskip 0.15em \mu $m and $h=15 \hskip 0.15em \mu $m. The patch dimension $l$ is varied within the range from $10 \hskip 0.15em  \mu $m to  $136  \hskip 0.15em \mu $m to cover a nearly full cycle of the phase response.\label{fig2}}
\end{figure}

The design of the terahertz reflectarray requires special attention on the choice of materials to satisfy micro-fabrication techniques. As shown in Fig.~\ref{fig2}, a resonant unit element for the terahertz reflectarray proposed here is made of three layers, from top to bottom: a gold patch, a polydimethylsiloxane (PDMS) substrate, and a platinum ground plane. Gold and platinum are good conductors, and are not subject to oxidization in air, whereas PDMS exhibits relatively low loss in the terahertz range. Different metals are chosen for the top and ground layer metallizations because of their selectivity for patterning as they react to different etching agents. If the same metal is used, permeation of the etchant through the PDMS will deteriorate the ground layer when patterning the top metallization.

In the simulation, the material parameters of the metals are obtained from a Drude model to determine the surface impedance $Z_{SR}$\cite{suf_impedance}, i.e.

\begin{equation}
\label{eq4}
Z_{SR} =\sqrt{\frac{j\omega\mu_{0}\mu_{r}}{\sigma_{R}+j\omega\epsilon_{0}}}, \hspace{0.4cm} \text{with} \hspace {0.1cm} \sigma_{R} = \frac{\sigma_{0}}{1+j\omega\tau},
\end{equation}
where $\sigma_{R}$ is the bulk complex conductivity of the metal at the considered frequency, $\sigma_{0}$ the DC-conductivity, $\tau = 1/\gamma_{p}$ the relaxation time, $\gamma_{p}$ the damping frequency,  $\mu_{0}$ the permeability of free space, $\mu_{r}$ the relative permeability, $\epsilon_{0}$ the permittivity of free space, $\omega=2\pi f$ the angular frequency, and $f$ the frequency of the incident wave. For gold and platinum, the corresponding parameters are $\sigma_{0,\text{Au}} = 4.10\times10^7$~S/m, $\sigma_{0,\text{Pt}} = 9.43\times10^6$~S/m, $\gamma_{p,\text{Au}} = 6.48\times10^{12}$~Hz, and $\gamma_{p,\text{Pt}} = 16.73\times10^{12} $~Hz. At the operation frequency $f = 1$~THz, the surface impedance of gold and platinum can be calculated as $Z_{SR,\text{Au}} = 0.287 + j\hspace{0.05cm}0.335$~$\Omega$ and $Z_{SR,\text{Pt}} = 0.628 + j\hspace{0.05cm} 0.667$~$\Omega$, respectively. The relative permittivity and loss tangent of PDMS are 2.35 and 0.03, respectively, as determined from measurement~ \cite{khodasevych2012elastomeric}.

\begin{figure}[b]
\centerline{\includegraphics[width=12cm]{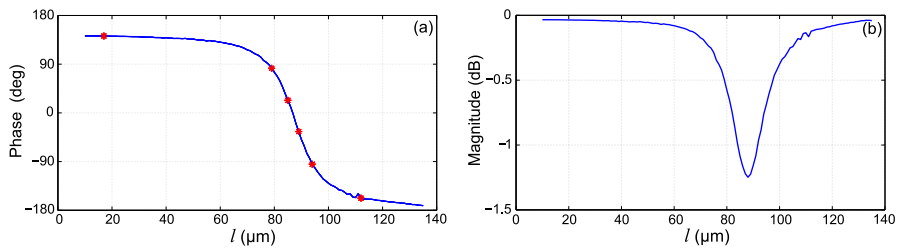}} \caption{Simulated reflection coefficient for a uniform 2D infinite patch array. Reflection phase response in degree (a) and reflection magnitude in dB (b) at 1~THz as a function of the patch size. The six points on the phase curve indicate the selected patch sizes to define a sub-array that fulfills one full cycle phase change. The roughness in the magnitude and phase curves is due to the limitation in the numerical accuracy. \label{fig3}}
\end{figure}

For the wave deflected from the surface of the gold patch, the local phase shift can be controlled by changing one or several parameters of the unit element, such as the size of the square gold patch $l$ or the thickness of the substrate $h$. By taking the design and fabrication feasibility into account, the side length $l$ of the gold patch is chosen as a variable, while the unit cell size and thickness of the PDMS substrate are fixed to $a=140 \hskip 0.15em \mu $m and $h=15 \hskip 0.15em \mu $m, respectively. The unit element is optimized at 1 THz by simulations using the Ansys HFSS commercial software with master-slave boundary conditions. When the length of the gold patch $l$ is varied within the range from  $10 \hskip 0.15em  \mu $m to  $136  \hskip 0.15em \mu $m, the magnitude and phase of the simulated reflection coefficient for a uniform 2D infinite patch array change as shown in Fig.~\ref{fig3}. The simulation results show that for the considered geometry and materials, the maximum range of the phase shift covers about $330^\circ$, which is close to a full cycle and sufficient for the intended operation of a reflectarray. In addition, low-loss reflection is observed for all of the investigated patch sizes with the highest loss of only about -1.2 dB on resonance.

\subsection{Design and simulations of the reflectarray}
\label{sec:Design and simulations of the reflectarray}

\begin{figure}[t]
\centerline{\includegraphics[width=10cm]{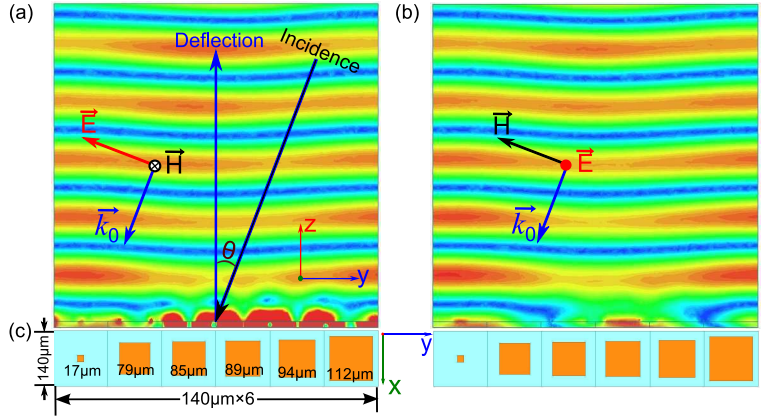}} \caption{Instantaneous scattered field from the reflectarray in TM and TE polarization at 1~THz. (a) Field distribution for the TM polarization. (b) Field distribution with the same structure and incident direction as in (a) but for the TE polarization. The incident wave is off normal with $\theta=21^\circ$. For the TM polarization, the $E$ field is in the $yz$ plane, and for the TE polarization, the $E$ field is in parallel with the $x$ axis. (c) Structure of one sub-array made of 6 patch elements depicted at the same scale as those in (a) and (b). \label{fig4}}
\end{figure}

Based on the relation between the phase change and the patch size shown in Fig.~\ref{fig3}, a reflectarray with off-specular reflection in one plane is designed. The progressive phase change $\Delta \phi$ is fixed at $60^\circ$. Therefore the number of elements for one linear sub-array amounts to 6 so that it covers one cycle of $360^\circ$. The unwrapped phase between the first and last elements of the sub-array also amounts to $60^\circ$, so that a periodic arrangement of the sub-array fulfills the desired deflection function. According to Eq.~\ref{eq3}, the deflection angle $\theta$ is calculated to be around $21^\circ$. As shown in Fig.~\ref{fig3}(a), the chosen 6 elements in the linear sub-array exhibit the phases decreasing from $142^\circ$ to -$158^\circ$ in the $60^\circ$ increment. This corresponds to increasing gold patch side lengths of $17\hspace{0.05cm}\mu\text{m}$, $79\hspace{0.05cm}\mu\text{m}$, $85\hspace{0.05cm}\mu\text{m}$, $89\hspace{0.05cm}\mu\text{m}$, $94\hspace{0.05cm}\mu\text{m}$, and $112\hspace{0.05cm}\mu\text{m}$. By the principle of reversibility of light, if the wave is incident with an angle of $21^{\circ}$ away from the normal, the direction of the deflected wave will be perpendicular to the surface of the reflectarray. Therefore, for convenience of observation, the incident wave is set with an angle of $21^{\circ}$ in the simulation with HFSS.

The numerically resolved instantaneous field distributions of the deflected wave for the TM and TE polarizations are shown in Fig.~\ref{fig4}(a) and (b), respectively. It is clear that the plane wave is deflected toward the normal direction, in close accordance with the theory. The field distribution for both the TM and TE polarizations is similar. A slight difference can be observed in the immediate proximity to the surface. This difference can be explained by the orthogonal mode field distributions under the patch elements for the two polarizations. Away from the surface, the slight deviation from a perfect plane wave is explained by the following effects. Firstly, inter-element coupling is different in uniform and nonuniform arrays. In the optimization of the single element, an infinite array with identical elements is considered, and the relation shown in Fig.~\ref{fig3} is obtained based on this assumption. In contrast, in the configuration of the sub-array, the dimensions of the neighboring elements vary in one direction, resulting in a different coupling behavior. Generally, smaller variations in the reflection phase of successive elements result in a flatter wave front, and therefore a better reflectarray performance. Secondly, the stronger attenuation of the patches near resonance, as illustrated by the magnitude curve in Fig.~\ref{fig3}(b), affects the uniformity of the deflected wave. Thirdly, the resolution of the patch size in the simulation has been limited to $1\hspace{0.05cm}\mu\text{m}$ to reflect the tolerance inherent to the fabrication process. This resolution limitation is relatively coarse and can cause significant error for the required phase response particularly around the resonance. It is clearly observed in Fig.~\ref{fig3}(a) that the phase response close to resonance is very sensitive to minor inaccuracies of the patch size. A less steep curve of the phase versus the patch size can decrease the sensitivity to the tolerances, however at the cost of a reduced range of available phases.

\section{Fabrication and measurement}

In order to validate the design, the reflectarray configuration shown in Fig.~\ref{fig4}(c) has been fabricated and measured. The details of the fabrication process and measurement system are given in this section.

\subsection{Fabrication}

The terahertz reflectarray antennas are fabricated using microfabrication and polymer processing techniques on $3"$ silicon substrates. The silicon (100) oriented substrates are cleaned in solvents (acetone and isopropyl alcohol) and dried using high purity compressed nitrogen. A $20\hspace{0.05cm}\text{nm}$ layer of titanium to serve as an adhesion promoter and a $200\hspace{0.05cm}\text{nm}$  thick layer of platinum for the ground plane are deposited from $99.99\%$ pure discs by electron beam evaporation at room temperature following pumpdown to a base pressure of $1\times10^{-7}$~Torr. PDMS, a silicone polymer prepared as a two-part mixture in a 1:10 ratio of hardener and pre-polymer, is spun on to the surface of the platinum coated wafers. This PDMS layer defines the dielectric in the reflectarray antenna. As the PDMS thickness is a critical parameter, the thickness dependence as a function of the spin speed at a fixed acceleration of $1,000 \hspace{0.05cm}\text{rpm}/\text{s}^2$ and duration of $30\hspace{0.05cm}\text{s}$ is experimentally determined. This is defined as an equation that presents a spin speed ($r$,~in~rpm) for a desired PDMS thickness ($h$,~in~$\mu\text{m}$) as:

\begin{equation}
\label{eq5}
r =	0.0001h^4 - 0.0328h^3 +3.9880h^2 - 238.460h +7926.4\;.
\end{equation}
For this work, to attain a $15\hspace{0.05cm}\mu\text{m}$ thick PDMS layer, a spin speed of 5,000 rpm is used. The spun on PDMS layer is cured at $72^\circ\text{C}$ for 1 hour. A $200\hspace{0.05cm}\text{nm}$ gold layer, with a $20\hspace{0.05cm}\text{nm}$ chromium adhesion layer, is then deposited by electron beam evaporation. These metal layers are patterned to define the antenna patches by photolithography and wet etching. The samples are then cleaned with solvents to strip residual photoresist in preparation for terahertz measurements.

\subsection{Measurement system}

\begin{figure}[b]
\centerline{\includegraphics[width=10cm]{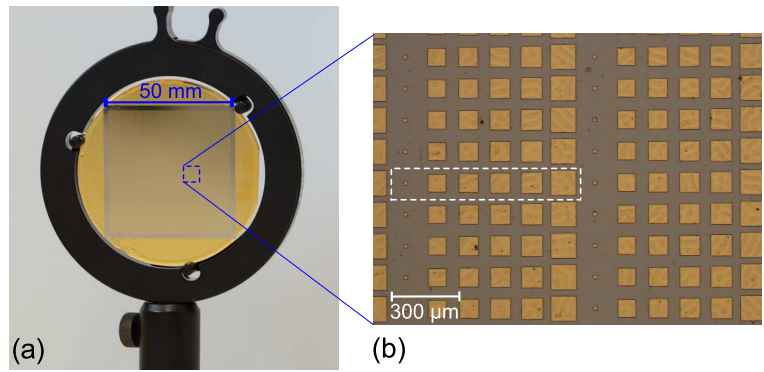}} \caption{Reflectarray prototype. (a) Photograph of the sample. (b) Microscopy image for a small part of the reflectarray. The dashed rectangle encloses one of the sub-arrays. \label{fig5}}
\end{figure}

\begin{figure}[htbp]
\centerline{\includegraphics[width=12cm]{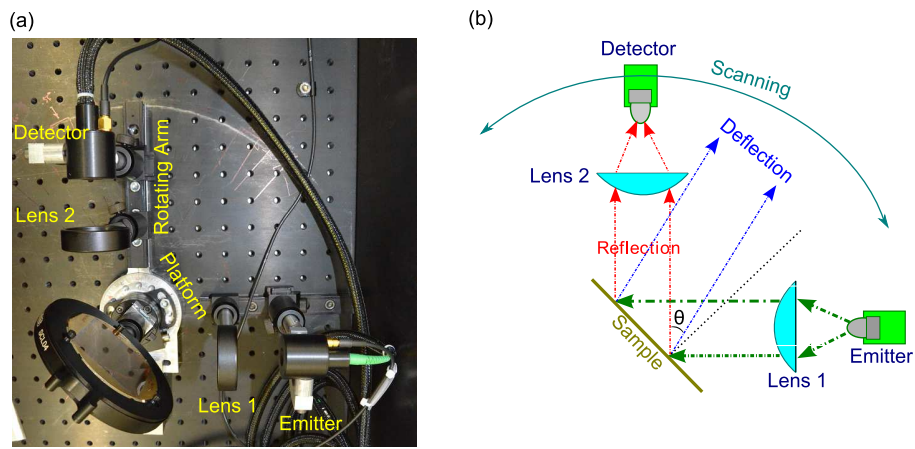}} \caption{Measurement system. (a) Photograph of the measurement system.  (b) Corresponding schematic. The beam from the emitter is collimated by Lens~1, and incident on the surface of the sample. Lens~2 concentrates the scattered beam into the detector. Lens~2 and the detector are fixed on an arm mounted on a rotating platform, allowing a wide angular range to be scanned.\label{fig6}}
\end{figure}

The sample shown in Fig.~\ref{fig5}(a) is made of $360\times360$ patch elements with periodic sub-array arrangement. The microscopy image shown in Fig.~\ref{fig5}(b) reveals the details of a small area of the sample. The THz-TDS measurement setup is shown in the photograph of Fig.~\ref{fig6}(a) with a corresponding schematic representation in Fig.~\ref{fig6}(b). A femtosecond optical pulse is guided by a fiber from a near-infrared laser source to the terahertz emitter. The generated broadband terahertz radiation is then guided from the emitter to the reflectarray via Lens~1 that collimates the divergent terahertz beam from the emitter. The parallel beam is either reflected or deflected, depending on the frequency, when it is incident on the surface of the reflectarray sample. The detection part of the system, comprising Lens~2 and the detector, is mounted on a rotatable arm for scanning the radiation in a wide angular range. On this arm, Lens~2 focuses the scattered radiation into the detector. All the reflectarray measurements are normalized by the free-space reference to remove any system dependency. For the reference, a gold-coated mirror substitutes the reflectarray, and the incident and reflection angles are set to 45$^\circ$. All measurements are performed under ambient temperature in dry atmospheric conditions.

\section{Results and discussion}

\subsection{Measured reflection and deflection spectra}
\label{sec:Measured reflection and deflection spectra}

\begin{figure}[htbp]
\centerline{\includegraphics[width=12cm]{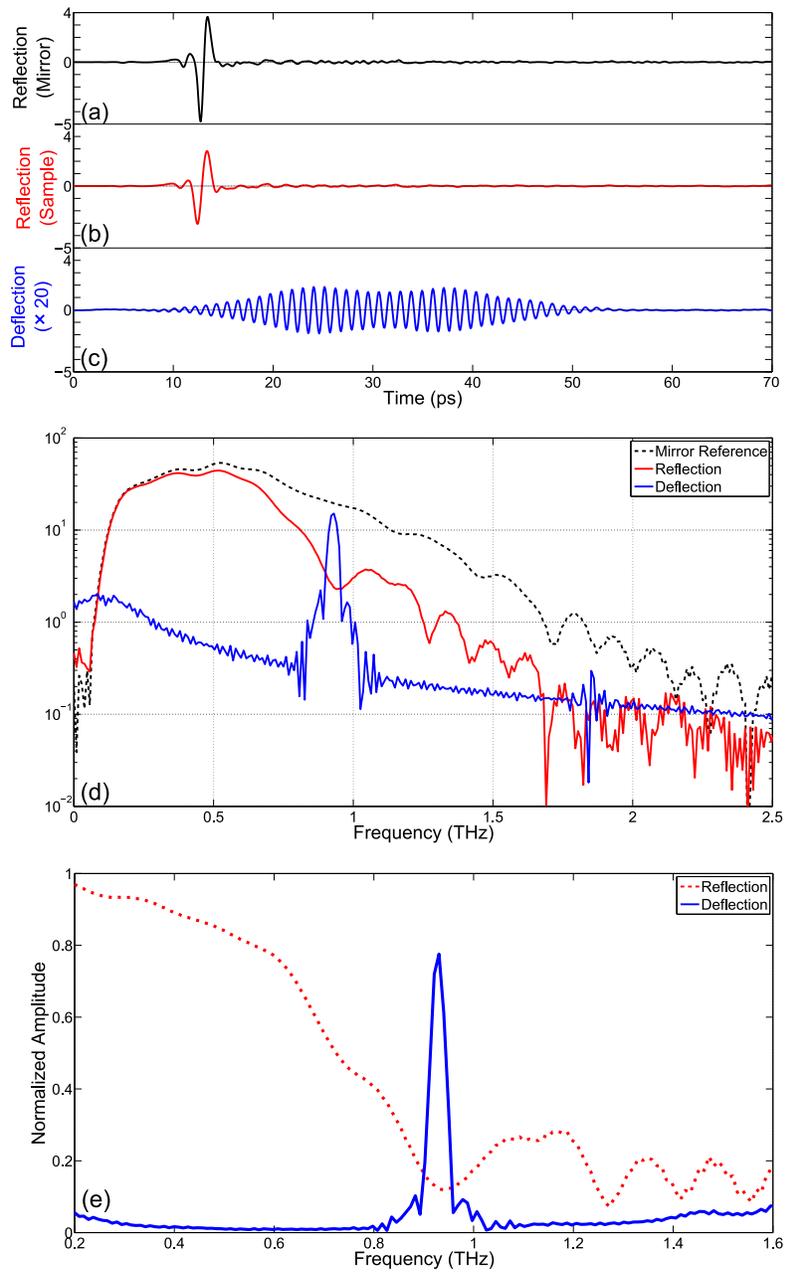}} \caption{Measured pulses and spectra in the TM polarization. (a) The reference pulse. (b) The specular reflection of the reflectarray sample. (c) The deflection of the reflectarray sample. (d) The spectra of the reference (black dashed line), the reflection (red solid line), and the deflection (blue solid line). (e) The normalized reflection (red dotted line) and deflection (blue solid line) amplitude.\label{fig7}}
\end{figure}

\begin{figure}[htbp]
\centerline{\includegraphics[width=12cm]{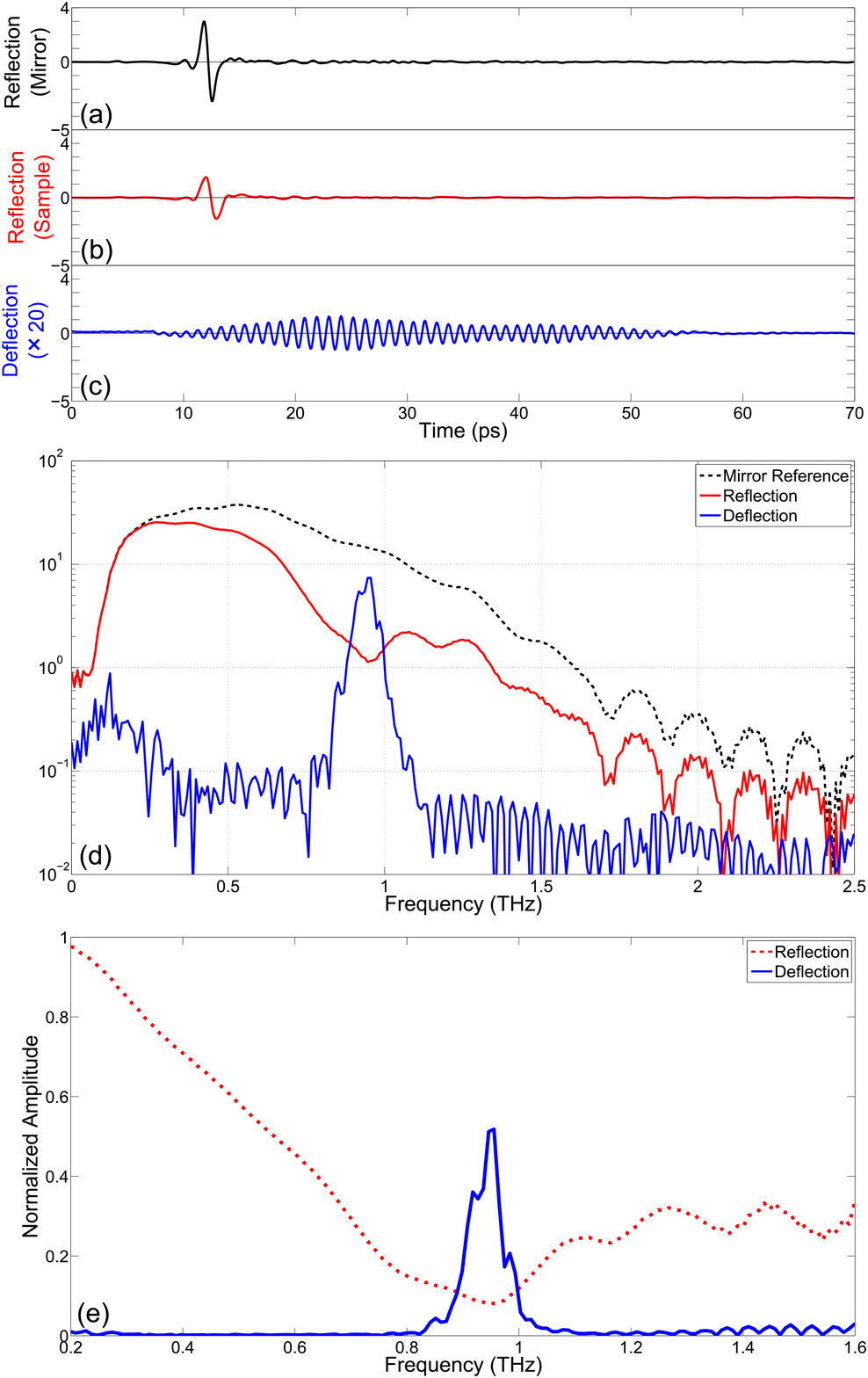}} \caption{Measured pulses and spectra in the TE polarization. (a) The reference pulse. (b) The specular reflection of the reflectarray sample. (c) The deflection of the reflectarray sample. (d) The spectra of the reference (black dashed line), the reflection (red solid line), and the deflection (blue solid line). (e) The normalized reflection (red dotted line) and deflection (blue solid line) amplitude.\label{fig8}}
\end{figure}

For the TM polarization, the reference pulse and its spectrum are presented in Fig.~\ref{fig7}(a) and Fig.~\ref{fig7}(d) (black dashed line), respectively. From 0.5 to 1.5~THz, the reference spectrum curve decreases smoothly without distinct absorption. The mirror is then replaced by the reflectarray sample to register the reflection in the specular direction. A strong reflection is detected, as shown by the pulse and corresponding spectrum in Fig.~\ref{fig7}(b) and (d), respectively. The reflection spectrum in  Fig.~\ref{fig7}(d) (red solid line) reveals an obvious notch around 0.93 THz. This means that considerable energy around this frequency is deflected off the direction of the specular reflection. The rotating arm is then moved to the expected angle of the deflection, and slightly adjusted for the maximal amplitude. The time-resolved deflection signal shown in Fig.~\ref{fig7}(c) exhibits an oscillation caused by the spectrally selective deflection of the reflectarray. This is confirmed in the deflection spectrum in Fig.~\ref{fig7}(d) (blue solid line), where a strong deflection peak appears at the frequency corresponding to the strongest notch in the reflection spectrum. Hence the measurement proves that the fabricated terahertz reflectarray has an ability to deflect the incident wave towards the predesigned direction. In order to estimate the performance of the reflectarray, the normalized reflection and deflection are calculated and shown in Fig.~\ref{fig7}(e), demonstrating that up to nearly $80\%$ of the incident amplitude is deflected around the operation frequency. It is worth noting that the sum of the reflection and deflection energy is less than unity at a wide frequency range. This missing energy is likely to be absorbed by the PDMS substrate or scattered into other directions. For the TE polarization, the results are given in Fig.~\ref{fig8}(a)-(e). The measurement results for both polarizations are similar. In both of the cases, the strong deflection is observed at around 0.93~THz.

\begin{figure}[htbp]
\centerline{\includegraphics[width=14cm]{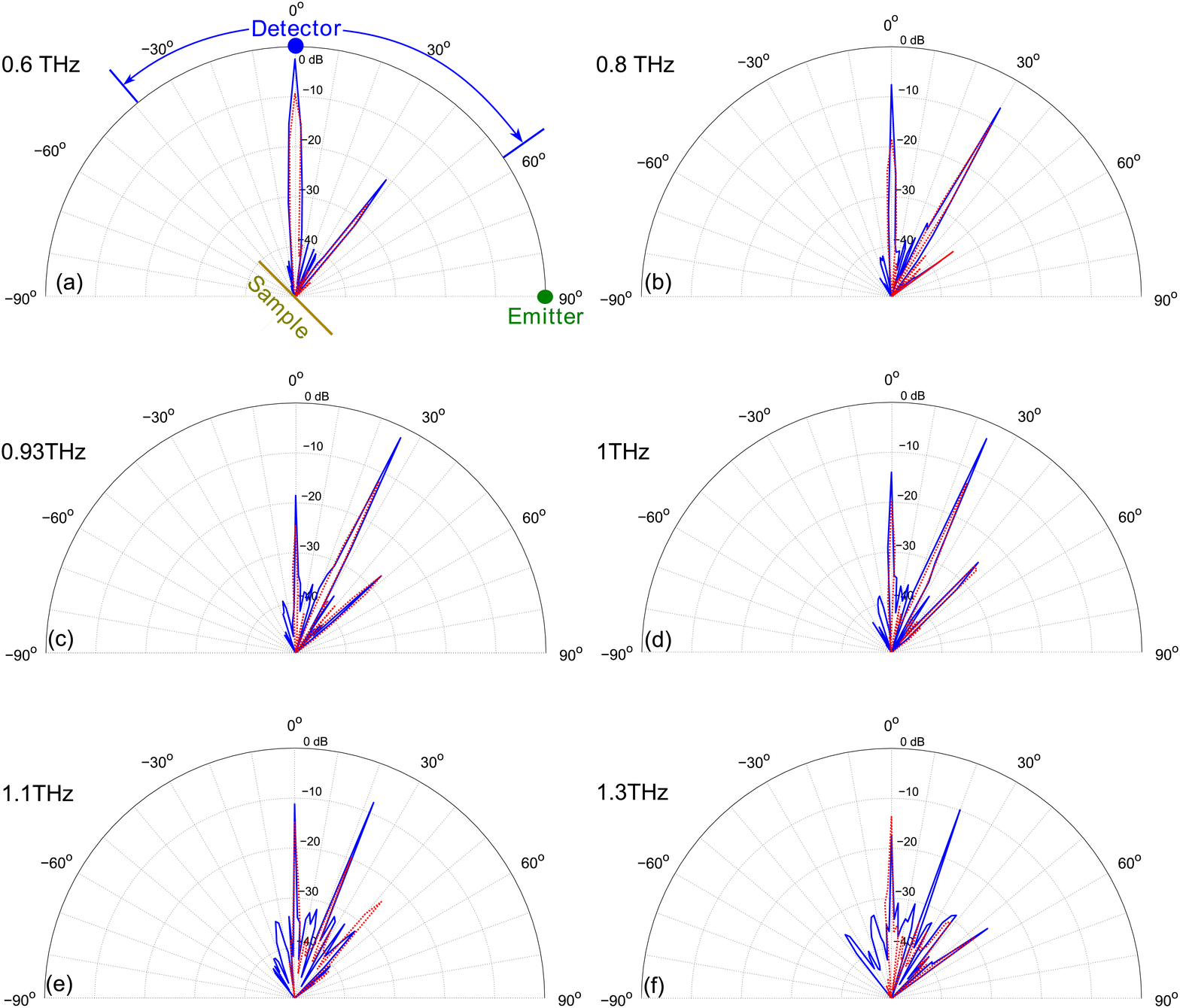}} \caption{Measured radiation pattern at different frequencies in TM polarization (blue solid line) and TE polarization (red dotted line) in logarithmic scale. The reflection and deflection always exist as Floquet modes although the intensity varies with frequencies. The closer the frequency is to 0.93 THz, the stronger deflection and the weaker reflection become, and vice versa. \label{fig9}}
\end{figure}

The angular radiation patterns of the reflectarray have been measured to characterize the spectral behavior of the reflected/deflected beams and side lobes. The radiation patterns are measured with an angular resolution of $2^\circ$ and are represented at different frequencies for both polarizations in Fig.~\ref{fig9}. At around 0.93~THz, as shown in Fig.~\ref{fig9}(c), the deflection is strongest, while on the other hand at 0.6~THz, the specular reflection is the strongest. At other frequencies, the patterns show a combination of lobes caused by the Floquet modes arising from the sub-array periodicity. Generally, the performance of the reflectarray for the TM and TE polarizations is similar.

\subsection{Discussion}

Despite a qualitatively satisfying demonstration of the reflectarray operation, there are some discrepancies between the simulated and measured results. The frequency for the maximum deflection is shifted from the designed frequency of 1~THz to the measured 0.93~THz. Meanwhile, the deflection angle also shifts from the expected $21^\circ$ to the measured $25^\circ$. Possible causes have been investigated and are described in the following.

\begin{figure}[htbp]
\centerline{\includegraphics[width=14cm]{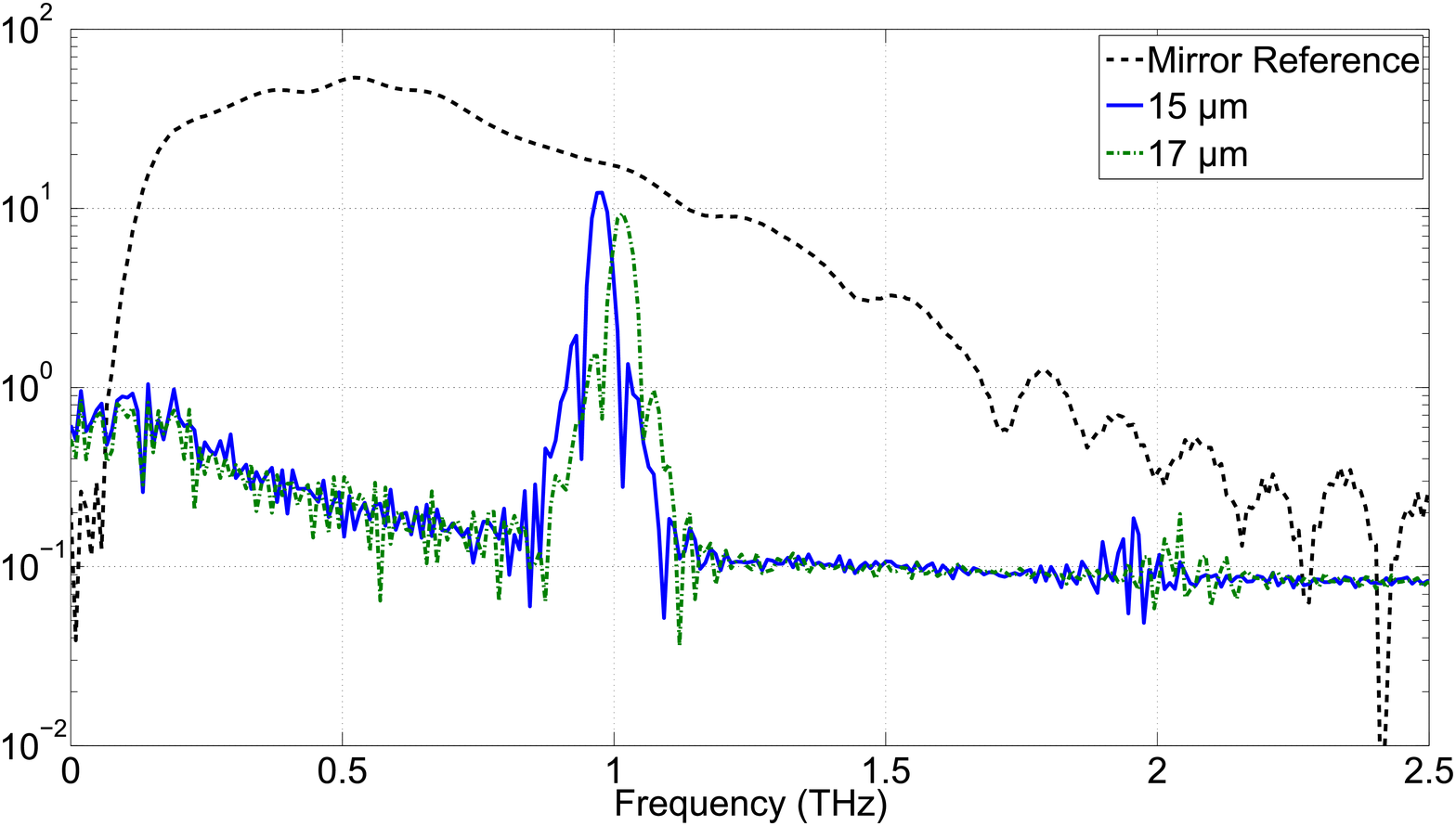}} \caption{Measured deflection spectra for two samples with different substrate thicknesses for the TM polarization. For the sample with the thickness of $15\hspace{0.05cm}\mu\text{m}$, the angle for maximum deflection is $25^\circ$, while for the $17\hspace{0.05cm}\mu\text{m}$ thick sample, the corresponding deflection angle is $26^\circ$. \label{fig10}}
\end{figure}

Fabrication tolerance is the main factor that gives rise to the frequency shift. In order to evaluate the effect from the tolerance, two samples with different substrate thicknesses ($15\hspace{0.05cm}\mu \text{m}$, $17 \hspace{0.05cm}\mu \text{m}$) have been measured. The resulting deflection spectra are shown in Fig.~\ref{fig10}. It is evident that a variation in the substrate thickness leads to a shift in the frequency for the maximum deflection. For the reflectarray with a substrate thickness of $15\hspace{0.05cm}\mu \text{m}$ investigated in Section~\ref{sec:Measured reflection and deflection spectra}, the measured frequency at 0.93~THz decreases from the designed frequency of 1~THz. Consequently, the deflection angle is increased according to the Floquet spatial mode (also called grating lobes) associated with the periodicity of sub-array structure. These Floquet modes cannot be neglected in arrays for which the inter-spacing of sub-array is larger than a half of wavelength. In the present case, the inter-spacing between adjacent sub-array is $840\hspace{0.05cm}\mu\text{m}$, or 2.8 times the wavelength at the frequency of operation.

In addition, the experiments are performed at an incident angle of $45^\circ$ rather than the normal incident angle used in the design step. This difference is another reason that causes the deflection angle to shift from the designed $21^\circ$ to the experimentally determined $25^\circ$. The dependence of the phase response curve on the incident angle has been investigated by Targonski~\emph{et~al.}~\cite{targonski1994analysis}. It was suggested that adopting the phase response of the normally incident wave for the oblique incident wave brings a new tolerance. Particularly, the difference between the practical performance and its theoretical expectation will become significant when the incident angle is larger than $40^\circ$.

Another reason for the discrepancy in the results is the mutual coupling between different patch elements that is not taken into account in the design process. In the fabricated reflectarray, the side length of nearby patches are not identical. Hence, as discussed in Section~\ref{sec:Design and simulations of the reflectarray},  the inter-element coupling becomes slightly different from that in the uniform patch array employed in the design process~\cite{milon2006analysis,milon2007surrounded}. The influence of the coupling will introduce inaccuracies in the phase response and therefore in the progressive phase change of the reflectarray elements. These inaccuracies eventually affect the deflection angle and also cause a difference in the results for the TE- and TM-polarized waves.

\section{Conclusion}

In this paper, terahertz reflectarrays with metallic patch elements have been proposed. The theoretical design taking into account fabrication tolerances has been verified through the simulations and experiments. The prototypes have been fabricated and the measurements have been carried out by using a THz-TDS system. The measurement in both TE and TM polarizations shows that a nonuniform reflectarray can efficiently deflect the terahertz waves towards a predesigned angle at a predefined frequency of operation. The possible factors for a small shift in the operation frequency and in the deflection angle have been investigated. It is suggested that the fabrication tolerance, the incident angle, and the mutual coupling effect should be taken into consideration in optimizing reflectarrays.

The proposed terahertz reflectarrays can become useful in various applications owing to their capability of manipulating terahertz beams with high efficiency yet low design and fabrication complexity. Their function is not limited to beam deflection, and can be extended to beam steering or shaping in various forms. In addition, active patch-element structure can be used to dynamically configure versatile arrays for advanced beamforming. In particular, the extension to active reflectarray systems promise the application in the area of short-range terahertz communications.

\section*{Acknowledgements}

Withawat Withayachumnankul, Madhu Bhaskaran, and Sharath Sriram acknowledge Australian Post-Doctoral Fellowships from the Australian Research Council (ARC) through Discovery Projects DP1095151, DP110100262, and DP1092717, respectively. Christophe Fumeaux acknowledges the ARC Future Fellowship funding scheme under FT100100585. The authors acknowledge technical assistance of Brandon Pullen, Longfang Zou, Jining Li, Hungyen Lin, and Henry Ho.

\end{document}